\documentstyle[twoside,fleqn,espcrc2,fps]{article}

\newcommand{\AmS}{{\protect\the\textfont2
  A\kern-.1667em\lower.5ex\hbox{M}\kern-.125emS}}
\begin{document}

\title{Simulating the vacuum properties of QCD with dynamical quarks}

\author{Xiang-Qian Luo 
\address{CCAST
     (World Laboratory),  P.O. Box 8730, Beijing 100080, China,\\
       Department of Physics,
       Zhongshan University, Guangzhou 510275, 
       China 
       (Mailing address)
       \\
       and Center for Computational Physics,
       Zhongshan University, Guangzhou 510275, China}}

\begin{abstract}
The vacuum properties of lattice QCD with staggered quarks are
investigated by an efficient simulation method. I present data for
the quark condensate with flavor number $N_f=0, ~ 1,
~ 2, ~ 3, ~ 4$ and many quark masses, including the vacuum energy
in the chiral limit. 
Obvious sea quark effects are observed in some parameter space.
I also describe a mechanism to understand this and a 
formula relating the chiral condensate 
and zero modes.
\end{abstract}

\maketitle


\section{INTRODUCTION}

Lattice QCD is a fundamental theory of non-perturbative strong interactions,
which should tell us everything about low energy hadronic physics. 
Here I would study the vacuum properties of full QCD
(unquenched theory). As is known, the most important issues are confinement,
chiral-symmetry breaking and sea quark effects \cite{Plenary}.
For the pure gluonic sector, there have been active
studies of the topology of the vacuum \cite{Plenary}.
For the fermionic sector, the interplay between
zero modes \cite{Plenary} and chiral-symmetry breaking
should be further investigated.

Earlier studies of this issue
employed quenched approximation, where 
sea quarks configurations are completely ignored.
Quenched Chiral Perturbation Theory (QCPTh)
predicts existence of some possible extra
chiral $\ln M_\pi $ terms \cite{QCPT}, which do not exist in full QCD. 
If this is the case, the quenched theory is wrong and will lead to
misbehavior of some quantities.

However, it is difficult in practice to visualize this misbehavior,
due to the ambiguity in chiral extrapolation.
To calculate 
$<{\bar \psi} \psi>\vert_{m=0}$,
a mass term $m {\bar \psi} \psi$ is conventionally introduced, 
and then extrapolation 
at some small set of $m$ has to be implemented
using some unjustified fitting function.

In \cite{MTC}, the Hybrid Monte Carlo (HMC) algorithm was used to
simulate unquenched configurations with 4 flavor Kogut-Susskind quarks.
This is a big step towards the full theory. In most simulations, however,
sea quarks (for the configurations) are usually distinguished 
from the valence quarks (for the propagators). By fixing $m_{sea}$, and varying
$m_{valance}$, the formula
\begin{eqnarray*}
\langle {\bar \psi} \psi (m_{valence},m_{sea},V) \rangle 
\end{eqnarray*}
\begin{eqnarray*}
= {\int [dU] e^{-S_g} det \Delta (m_{sea})
\sum_j^{3V} 
{2m_{valence} \over \lambda_j^2 +m_{valence}^2} 
\over \int [dU] e^{-S_g}det \Delta (m_{sea})}
\label{ps}
\end{eqnarray*}
is used and then extrapolated 
to $m_{valence}=0$. Unfortunately, this is not theoretically
consistent. 

In the following sections, I would describe
a possibly better solution to these problems.

\section{ALGORITHM}

HMC is not efficient
for investigating the chiral properties. 
There exists an efficient algorithm for investigating the
vacuum properties of lattice gauge theory with dynamical fermions,
named the microcanonical fermionic average method \cite{MFA}.
This algorithm is based on the definition of an
effective fermionic action

\begin{eqnarray*}
S_{eff}^F(E,m,N_f)=
\end{eqnarray*}
\begin{eqnarray}
- \ln  {\int [dU] [det \Delta(m, U)]^{N_f/4}
\delta(S_g (U) - 6VE)
\over 
\int [dU]
\delta(S_g (U) - 6VE)},
\label{seff}
\end{eqnarray}
where $E$ is the gauge energy. The thermodynamical properties
can be obtained from the derivatives of the partition function 
\begin{eqnarray}
{\cal Z} = \int dE e^{-S_{eff}(E,m,N_f,\beta)}.
\end{eqnarray}
Here $S_{eff}$ is the full effective action related to
$S_{eff}^F$, $E$ and density of states $n(E)$ by 
\begin{eqnarray*}
S_{eff}(E,m,N_f,\beta)= - \ln n(E) 
\end{eqnarray*}
\begin{eqnarray}
- 6 \beta VE + S^{F}_{eff}(E,m,N_f).
\label{full}
\end{eqnarray}

This method has been tested in the Schwinger model \cite{QED2},
applied to QED in 3 dimensions \cite{QED3}, 
QCD in 3 dimensions \cite{QCD3}, QED in 4 dimensions \cite{QED4}, 
and generalized
to the fermion-gauge-scalar models \cite{FGS}.
For $m \not=0$, it has been cross-checked
with HMC and the data are in good agreement. 
(Fairly speaking, it is more convenient to use HMC to calculate the spectrum
and matrix elements in the non-chiral regime).
One prominent advantage is that this algorithm might be still valid in the chiral
limit. It can also be carried out 
on small computers or workstations. 

The results to be presented will be the first application
to QCD in 4 dimensions.

\section{RESULTS}

I describe only the $6^4$ lattice data, while those on larger lattices
will be given elsewhere.

\begin{figure}[htb]
\fpsxsize=6.2cm
\hspace{6mm}
\def\fpsangle{270}
\fpsbox[70 90 579 760]{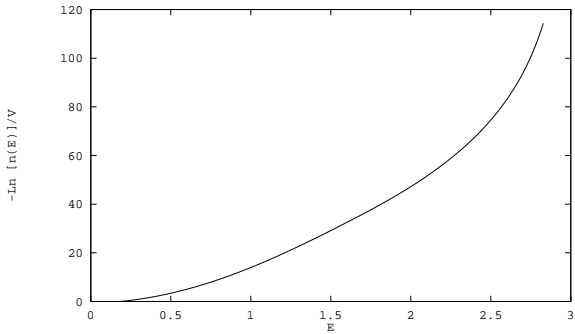}
\vspace{-20mm}
\caption{$- \ln [n(E)]/V$ as a function of $E$}
\label{fig1}
\end{figure}

\begin{figure}[htb]
\fpsxsize=6.2cm
\hspace{6mm}
\def\fpsangle{270}
\vspace{-2mm}
\fpsbox[70 90 579 760]{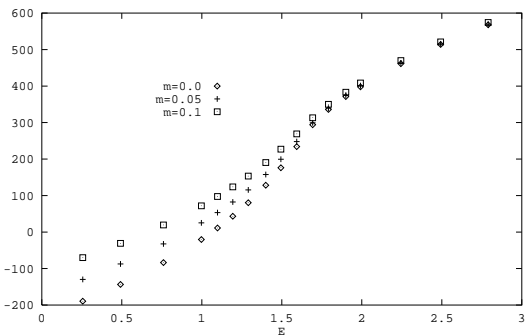}
\vspace{-20mm}
\caption{$-S^F_{eff}$ as a function of $E$ for $N_f=2$.}
\label{fig2}
\end{figure}

\begin{figure}[htb]
\fpsxsize=6.2cm
\hspace{6mm}
\def\fpsangle{270}
\vspace{-5mm}
\fpsbox[70 90 579 760]{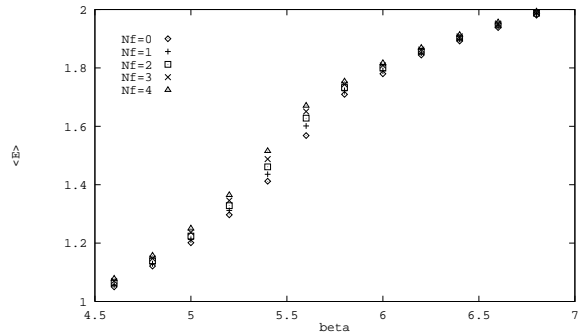}
\vspace{-20mm}
\caption{The mean plaquette energy $<E>$ as a function 
of $\beta$ in the chiral limit $m=0$.}
\label{fig3}
\end{figure}

\begin{figure}[htb]
\fpsxsize=6.2cm
\hspace{6mm}
\def\fpsangle{270}
\fpsbox[70 90 579 760]{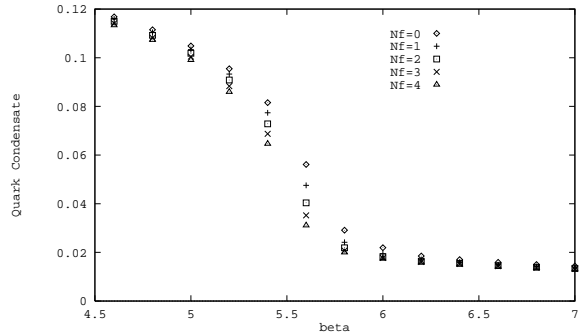}
\vspace{-20mm}
\caption{$<{\bar \psi} \psi>$ as a function of $\beta$ for $m=0.05$.}
\label{fig4}
\end{figure}

$n(E)$ in (\ref{full}) can be directly evaluated by numerically
integrating the quenched SU(3) data: 
\begin{eqnarray}
- {\ln n(E) \over V} =6 \int_{0}^{E} dE' \beta (E',N_f=0)+ const.,
\end{eqnarray}
as shown in Fig . \ref{fig1}. Fig. \ref{fig2} only shows the effective fermionic action as
a function of $E$. The shapes of the curves for other
$N_f$ are quite similar, but the scales are quite different. We have approximately
$S^F_{eff} \propto N_f$. Using the saddle point technique,  
the mean plaquette energy $<E>$ as a function of
$\beta$
is found, as shown in
Fig. \ref{fig3} for $m=0$ and different $N_f$. 
To my knowledge, $<E>\vert_{m=0}$  in QCD 
has never been obtained before. As seen in this figure, 
$<E>$ increases with $N_f$. The sea quark effects become most
important around $\beta \approx 5.5$, which corresponds to $<E> \approx 1.5$.
We can understand this phenomenon by looking at Fig. \ref{fig2},
from which one can observe maximum slope around $E \approx 1.5$.
The lighter the quark mass,
the bigger the slope is.
Therefore, we have a mechanism: the sea quark effects  
reach their maximum where the effective fermionic action
versus the gauge energy has a maximum slope. 
In Fig. \ref{fig4}, the results for $<{\bar \psi} \psi>$ at $m=0.05$ for
different $N_f$ are presented. Again, one sees $<{\bar \psi} \psi>$
decreases with $N_f$ and the sea quark effects
become dominant in the same region. Detailed analysis will be presented
soon.

\section{OUTLOOKS}

The estimate for $<{\bar \psi} \psi>\vert_{m=0}$ remains to be done.
An improved method \cite{Bar1}
(also \cite{VS}) uses
$\langle {\bar \psi} \psi \rangle =3 \pi \lim_{\lambda \to 0} \rho (\lambda)$.
For a finite $V$, the eigenvalue density
$\rho (0)=0$ due the absence of exact zero mode.
Again, some assumed fitting function has to be used.

In \cite{PDF}, Azcoiti, Laliena and I 
discovered that the chiral condensate
can be directly obtained by analyzing its probability distribution
function, constructed from the eigenvalues. We
derived 
\begin{eqnarray}
<{\bar \psi} \psi>\vert_{m=0} = {a_i \over V}  <{1 \over \lambda_i}>.
\end{eqnarray}
Here $a_i$ is the $ith$ zero of the Bessel function $J_0$. 
Only a few smallest eigenvalues are necessary for this calculation. 
It is worth mentioning that no $\lambda$ or $m$ extrapolation should be done.
The only thing to do is the finite size analysis. One can also understand apparently
the relation between zero modes and chiral-symmetry breaking. If chiral symmetry
is spontaneously broken, the eigenvalues $\lambda_i$
should behave as $a_i/V$ for large volume.
We have checked \cite{PDF} it in the Schwinger model,
where exact results are available. I do believe the method is promising for QCD.

This work was supported by National Natural 
Science Foundation under grant Nos.
19605009, 19677205,19710210523 (international exchange $\&$ collaboration), 
and by National Education Committee under grant
no. jiao-wai-si-liu[1996]644.
I thank the Lattice 97 organizers for support and assistance.
Discussions with V.Azcoiti, T.Chiu, G.DiCarlo, A.Galante, A.Grillo, S.Guo, 
H. Kr\"oger, V.Laliena, L.Lin, J.Liu, K.Liu, D.Sch\"utte and 
J.Wu are appreciated.

\end{document}